\documentclass[11pt]{amsart}
\usepackage{amssymb}
\usepackage{amsfonts}
\usepackage{amscd}
\usepackage{amsmath}
\usepackage{mathrsfs}
\usepackage[margin=1.2in, centering]{geometry}
\usepackage{graphicx}
\usepackage[shortlabels]{enumitem}
\numberwithin{equation}{section}

\newtheorem{theorem}{Theorem}[section]

\def \mcb {{\mathcal B}}
\def \mcc {{\mathcal C}}

\def \mcm {{\mathcal M}}

\def \mcr {{\mathcal R}}

\def \mcv {{\mathcal V}}
\def \mcw {{\mathcal W}}

\def \mbr {{\mathbb R}}
\def \mbs {{\mathbb S}}

\def \loc {\operatorname{loc}}  

\def \supp {\operatorname{supp}}

\def \WF {\operatorname{WF}} 
 
\def \sym{\operatorname{Sym}}

\def \re {\operatorname{Re}}

\def \beqq {\begin{equation}}
\def \eeqq {\end{equation}}
\def \bpf {\begin{proof}}
\def \epf {\end{proof}}
\def \beq {\begin{equation*}}
\def \eeq {\end{equation*}}

\def \eps {\epsilon}   
\def \la {\lambda}   
\def \La {\Lambda}     
\def \lap {\Delta}
\def \p {\partial}
\def \ha {\frac{1}{2}}    
\def \tf {\texttt{f}}
\def \th {\texttt{h}}
\def \tv {\texttt{v}}
\def \tw {\texttt{w}}

\begin{document}
\title[]{On finding gravitational waves from anisotropies of the cosmic microwave background}
\author{Yiran Wang}
\address{Yiran Wang
\newline
\indent Department of Mathematics, Emory University
\newline
\indent 400 Dowman Drive, Atlanta, GA, 30322, U.S.A. }
\email{yiran.wang@emory.edu}

\begin{abstract} 
The integrated Sachs-Wolfe (ISW) effect describes how photons are gravitationally redshifted, producing anisotropies in the Cosmic Microwave Background. We study the inverse problem and show that primordial gravitational perturbations, in particular their polarizations in the transversally traceless (TT) gauge can be identified from the local observation of the ISW effect.  
\end{abstract}
\date{\today. This work is partly supported by NSF grant DMS-2508559.}
\maketitle

\section{Introduction}  
\subsection{The physical problem}
Primordial gravitational waves (PGWs) generated in the beginning of Universe are of great interest in cosmology, see for example \cite{Dod, KDS, Tho}. However, their  detection is very challenging. As stated in \cite{KDS}, ``Direct detectors, such as the Laser Interferometer Gravitational-wave Observatory detector, are being designed to measure the local space-time distortion.... (PGWs) will involve waves today whose wavelengths will extend all the way up to our present cosmological horizon and that are likely to be well beyond the reach of any direct detectors for the foreseeable future." Despite the difficulties in direct observation, theoretical studies have shown that PGWs will leave signatures in the Cosmic Microwave Background (CMB). In particular, photons are polarized due to Thompson scattering. The polarization modes have been successfully detected, however separating the B-mode associated with gravitational waves (tensor perturbations) is a non-trivial task. To this date, the approach has not succeeded yet, and the identification of PGWs remains tantalizing. 

In this work, we consider another possibility of searching for PGWs from CMB anisotropies instead of polarizations. It is known since the work of Sachs and Wolfe \cite{SaWo} that PGWs produce CMB anisotropies, and recent studies (e.g., \cite{LAS}) indicate that some of the signatures could be on the observable level. The purpose of this work is to analyze the effects and show that tensor perturbations can be identified via a tomography approach.  

We start with the description of the physical problem. Let $(\mcm, g_0)$ be the Friedman-Lema\^itre-Robertson-Walker (FLRW) cosmological model, where 
\beqq\label{eq-uni0}
\mcm = (0, \infty)\times \mbr^3, \quad g_0(x) = -ds^2 + a(s)^2 dy^2. 
\eeqq
Here, $x = (s, y), s > 0, y\in \mbr^3$ is the coordinate for $\mcm$, and $a(s) > 0$ is smooth. In this model, the Universe starts  from a Big Bang at $s = 0$ and inflates.  The factor $a(s)$ reflects the rate of expansion. For example, when the Universe was very young and dominated by radiation, the factor $a(s)\approx s^{1/2}.$ At later time, when matter became to dominate, the factor $a(s)\approx s^{2/3}$. 
We consider the actual Universe $(\mcm, g)$ to be a metric perturbation of $g_0$. Within Einstein's theory of General Relativity, we assume that $g$ satisfies Einstein's equations. 

Next, consider the photon distribution in $(\mcm, g)$.  Let $0< s_0< s_1 <\infty$. We let $\mcm_0 = \{s_0\}\times\mbr^3$ be the ``surface of last scattering", where photons are formed and start to travel freely in $\mcm$.  Then we observe the photons on $\mcm_1 = \{s_1\}\times \mbr^3$. By ignoring photon interactions, the trajectories of photons can be described by null-geodesics in $(\mcm, g)$. Let $\gamma(\tau), \tau \in \mbr$ be a null geodesic from $\mcm_0$ to $\mcm_1$ in $(\mcm, g)$. Here, we assume for simplicity that $g$ is globally hyperbolic so that $\mcm_0, \mcm_1$ are Cauchy surfaces and any null geodesic intersects them only once.  Let $\Xi = \p_s$ be a time-like vector field whose flow line represents the observer. The initial energy of the photon observed by  $\Xi$ is $E_0 = g(\dot \gamma(0), \Xi).$ The energy received by the observer at $\mcm_1$ is
$E = g(\dot \gamma(\tau_1), \Xi). $  The redshift is defined by  
\beqq
R = \frac{E_0 - E}{E}. 
\eeqq
The Sachs-Wolfe effect is essentially the first order linearization of $R$, see \cite{SaWo, Dod, Dur}.  Suppose that the metric perturbation $g$ is a one parameter family 
\beqq\label{eq-geps0}
g_\eps = g_0 + \eps g_1 + \eps^2 g_2 + \cdots
\eeqq
and that the map $\eps\rightarrow g_\eps$ is differentiable from $(-\eps_0, \eps_0)$ to $C^2(\mcm; \sym^2)$ for some $\eps_0 > 0$ where $\sym^2$ denotes the bundle of symmetric covariant two tensors on $\mcm.$     
Let $R_\eps$ be the corresponding redshift for $g_\eps$.  It is proved in \cite[Theorem 4.2]{LOSU} (see also \cite[equation (39)]{SaWo}) that
\beqq\label{eq-tomo}
\p_\eps R_\eps|_{\eps = 0} = \frac{1}{2 a(s_0)}X (a^2 L_{a \Xi} (a^{-2}  g_1)). 
\eeqq
Here,  $L_\bullet$ denotes the Lie derivative for a vector field $\bullet$, and $X$ is an integral transform involving the integration along null geodesics on $(\mcm, g_0)$. More precisely, let $\gamma_0(\tau)$ be a null geodesic  for $g_0$ such that $\gamma_0(0)\in \mcm_0$ and $\gamma_0(\tau_0)\in \mcm_1$. For a symmetric two tensor field $f$, $Xf$ is defined by  
\beqq\label{eq-lray}
X f (\gamma_0) = \int_0^{\tau_0} \sum_{i, j = 0}^3  f_{ij}(\gamma_0(\tau))\dot \gamma_0^i(\tau) \dot \gamma_0^j(\tau) d\tau,
\eeqq
when the integral makes sense. The term in \eqref{eq-tomo} is the (integrated) Sachs-Wolfe (ISW) effect. We remark that even though ISW cannot be directly read off from the CMB data, there are many studies on extracting ISW from the CMB by combining it with other survey data, see \cite{MaDo, Sha1} for example.   

We study the inverse problem of recovering $g_1$ from the ISW \eqref{eq-tomo}. When the perturbation is of scalar type, that is $g_1 = \Phi dt^2 + \Phi dx^2$ for some scalar function $\Phi$, the problem is studied in \cite{ChWa, VaWa}, see also \cite{Wan2} for a survey of recent developments in cosmological X-ray tomography. Very little is known for the tensor problem. In this work, we prove results on the recovery of transversally traceless (TT)  gravitational waves, which is most relevant for analyzing the CMB polarizations, see for example \cite{HuWh, Kos}.

\subsection{The main results}\label{sec-main}
\begin{figure}[t] 
\centering
\includegraphics[scale = 0.52]{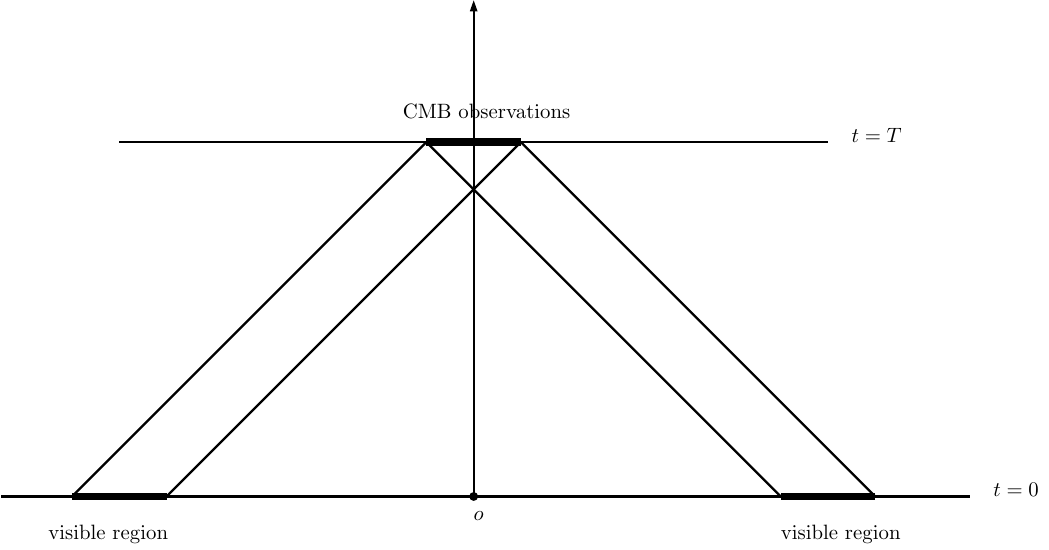}  
\includegraphics[scale = 0.6]{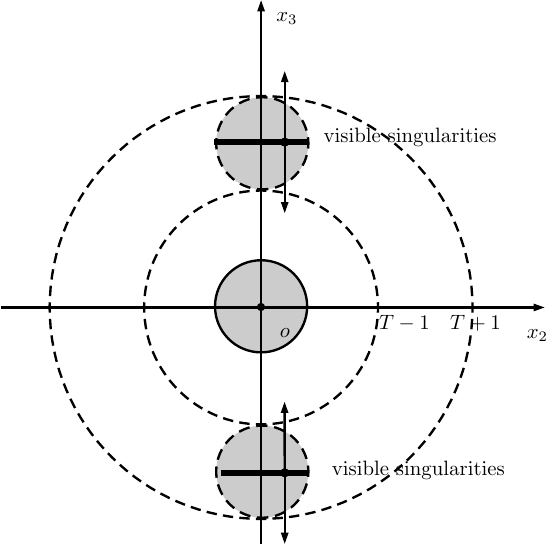}
\caption{Left figure: Illustration of Theorem \ref{thm-main0}. The picture shows the projection of $\mcm$ to $t-x_1$ plane. The visible region corresponds to $\mcr_{T-1, T+1}$. Right figure: Illustration of Theorem \ref{thm-main}. The picture shows the projected view of $\mcm$ to $t = 0$ on the $x_2-x_3$ plane. The ring bounded by two dashed circles is $\mcr_{T-1, T+1}$. The shaded disk in the middle is $\mcb_1$ representing the projection of the set on $t = T$ where the ISW is observed. The other two shaded disks are $\mcr^\pm$ where singularities of the initial data can be recovered.}
\label{fig-setup} 
\end{figure}

The metric in \eqref{eq-uni0} is conformal to the Minkowski metric. The conformal factor does not affect our analysis, so for simplicity, we consider the Minkowski spacetime as the background Universe. Let $\mcm = (0, T)\times \mbr^3$ and $(t, x), t\in (0, T), x\in \mbr^3$ be the local coordinates. Let  $g = -dt^2 + \sum_{i = 1}^3dx_i^2$ be the Minkowski metric on $\mcm$.   For $t\in \mbr,$ we denote $\mcm_t = \{t\}\times \mbr^3$.  Let $\mcb_1  = \{x\in \mbr^3: |x| < 1\}$ be the unit ball on $\mbr^3$. We consider null geodesics from $\mcm_0$ that meet $\mcb_1$ on $\mcm_T$. See Figure \ref{fig-setup}. This models the local observation of CMB near the Earth. It is convenient to parametrize the null geodesics using $x\in \mcb_1$ and $v\in \mbs^2$ as  
\beqq\label{eq-geo}
\gamma_{x, v}(t) = (t, x + tv - Tv), \quad t\in \mbr.
\eeqq
For $t\in (0, T)$, $\gamma_{x, v}(t)$ is contained in $\mcm$. Note that the set of such geodesics can be identified with the set $\mcc = \mcb_1 \times \mbs^2.$ For a symmetric two tensor field $u$ in $\mcm$, we consider the light ray transform
\beqq\label{eq-light}
 X u(x, v) = \int_0^T \sum_{i, j = 0}^3 u_{ij}(t, x + t v - Tv)\theta^i\theta^j ds, \quad (x, v)\in \mcc, 
\eeqq
where $\theta = (1, v)$. Furthermore, we consider $u$  satisfying  the Cauchy problem for the tensorial wave equation 
\beqq\label{eq-wave}
\begin{gathered}
\square u(t, x) = 0, \quad (t, x)\in \mcm \\
u(0, x) = f(x), \quad \p_t u(0, x) =  h(x). 
\end{gathered}
\eeqq 
It is well-known that \eqref{eq-wave} is the linearized vacuum Einstein equations in the Lorentz gauge at the Minkowski background, see \cite[Chapter 18]{MTW}. This is sufficient for the investigation of the linearized CMB theory.   

We study the recovery of $f, h$ from $Xu$ on $\mcc$. We first observe that this is not always possible. Let $\mcb_{1+T} = \{x\in \mbr^3: |x| <1+ T\}$ on $\mcm_0$. If $f, h$ are supported outside of $\mcb_{1+T}$, by the finite speed of propagation of \eqref{eq-wave}, we see that $\supp u$ does not meet any light rays that intersect $\mcb_1$ at $t = T$ so $Xu = 0$ on $\mcc$. Thus the best we can hope for is to recover $f, h$ in $\mcb_{1+T}.$ For $b>a>0$, we set 
\beqq\label{eq-ring}
\mcr_{a, b} = \{x\in \mbr^3: a< |x|< b\}. 
\eeqq
We use the convention that if $a\leq 0$ then $\mcr_{a, b} = \mcb_{b}$. Our main results concerns the recovery of information in $\mcr_{T-1, T+1}$. 

In the study of CMB polarizations, it is typical to consider transversally traceless (TT) gravitational waves, see e.g., \cite[Section 2.4]{HuWh}. 
We follow the presentation in \cite[Chapter 35]{MTW}. We start with the monochromatic plane wave solutions of \eqref{eq-wave} of the form 
\beqq\label{eq-plane}
u_{ij} = \re(A_{ij}e^{-\imath \zeta \cdot x}), \quad i, j = 0, 1, 2, 3, 
\eeqq
where $\imath= \sqrt{-1}$, $\zeta \in \mbr^4$ is the wave vector, and $A\in M_{4\times 4}$ is the amplitude. In particular, 
they satisfy  $\sum_{i = 0}^3\zeta_i \zeta^i = 0$ and $\sum_{j = 0}^3 A_{ij}\zeta^j = 0$. The dynamic degree of freedom for a gravitational field in the linearized theory is two, which can be revealed by imposing proper gauge conditions. First, consider the global Lorentz frame associated with the $4$-velocity $\Xi = (1, 0, 0, 0)$. Then we consider symmetric two tensor $u$ satisfying 
\begin{enumerate}[(i)]
\item $u_{i0 = 0}, i = 0, 1, 2, 3$, meaning only the spatial components of $u$ are possibly non-zero. 
\item $\sum_{j = 1}^3 u_{ij, j} = 0, i = 1, 2, 3$,  meaning the spatial components of $u$ are divergence free. 
\item $\sum_{j = 1}^3 u_{jj} = 0,$ meaning the spatial components of $u$ are trace-free.  
\end{enumerate}
Such tensors are called TT tensors. It is known that for a specific global Lorentz frame of the linear theory, one can impose gauge conditions (i)-(iii) to describe a specific gravitational wave, see \cite[Chapter 35]{MTW}. In TT gauge, the plane wave solution \eqref{eq-plane} can be written in the form 
\beq
u(t, x) = 
\begin{pmatrix}
0 & 0 & 0 & 0 \\
0 & A_+ & A_\times & 0 \\
0 & A_\times & -A_+ & 0 \\
0 & 0 & 0 & 0 
\end{pmatrix} 
\cdot \re(e^{-\imath \omega (t - x_3)}), 
\eeq
where $A_+, A_\times$ are constants representing the two independent mode of polarizations, and $\omega = \zeta^0$ is the frequency of the wave. We see that $u$ is a plane wave traveling along the $x_3$ axis and the oscillation is transversal to the direction of the  propagation. More generally, we can consider superposition of such plane waves. Note that the gauge conditions (i)-(iii) are linear and preserved under \eqref{eq-wave}. We consider initial conditions 
\beqq\label{eq-fh}
f(x) = 
\begin{pmatrix}
0 & 0 & 0 & 0 \\
0 & f_+(x_3) & f_\times(x_3) & 0 \\
0 & f_\times(x_3) & -f_+(x_3) & 0 \\
0 & 0 & 0 & 0 
\end{pmatrix} , \quad 
h(x) = 
\begin{pmatrix}
0 & 0 & 0 & 0 \\
0 & h_+(x_3) & h_\times(x_3) & 0 \\
0 & h_\times(x_3) & -h_+(x_3) & 0 \\
0 & 0 & 0 & 0 
\end{pmatrix}. 
\eeqq
Then the solution $u$ of \eqref{eq-wave} is a TT wave traveling along the $x_3$ axis. Note that $f, h$ and $u$ are not compactly supported. This is not a problem  because only part of $f, h$ can be recovered from the partial observation of CMB. 

Our first result concerns the unique determination of $f, h.$ 
\begin{theorem}\label{thm-main0}
Let $u$ be a TT wave satisfying \eqref{eq-wave} with Cauchy data $f \in H^s_{\loc}(\mbr^3), h \in H^{s-1}_{\loc}(\mbr^3), s>2$ of the form \eqref{eq-fh}. 
Then $f, h$ on $\mcr_{T-1, T+1} $are uniquely determined by $Xu$ on $\mcc$, namely if $Xu = 0$ on $\mcc$, then $f = h = 0$. 
\end{theorem}

Next, we consider  ``inverting"  $X$ to recover information of $f, h$. The proof of Theorem \ref{thm-main0} relies on the analytic continuation which provides little information in this regard. Our next result gives the inversion operator in the microlocal sense that tells what singularities of $f, h$ can be recovered from $Xu$. Let $e_3 = (0, 0, 1)$ and $\mcr^\pm  = \mcb_1 \pm Te_3$ be the translation of $\mcb_1$ along the $e_3$ direction. Then we define    
\beqq\label{eq-wpm}
\begin{gathered}
\mcw^+  = \{(x, \xi)\in T^*\mbr^3\backslash 0: x\in \mcr^+, \xi =  \la e_3, \la >0 \text{ or } x\in \mcr^-, \xi = - \la e_3, \la >0 \}, \\
\mcw^-  = \{(x, \xi)\in T^*\mbr^3\backslash 0: x\in \mcr^+, \xi = - \la e_3, \la > 0 \text{ or } x\in \mcr^-, \xi =  \la e_3, \la > 0 \}. 
\end{gathered}
\eeqq 
\begin{theorem}\label{thm-main}
Let $u$ be a TT wave satisfying \eqref{eq-wave} with Cauchy data $f \in H^s_{\loc}(\mbr^3), h \in H^{s-1}_{\loc}(\mbr^3), s>2$ of the form \eqref{eq-fh}. Consider the re-parametrized Cauchy data 
\beq
\begin{gathered}
\tf_+ =  f_+ + \lap^{-1/2}  h_+, \quad \tf_\times =  f_\times +   \lap^{-1/2}  h_\times, \\
\th_+ =  f_+ -  \lap^{-1/2} h_+, \quad \th_\times =  f_\times -  \lap^{-1/2} h_\times. \\
\end{gathered}
\eeq
Then there exists operators $H_\times, H_+$ given in \eqref{eq-htimes} and \eqref{eq-hplus} such that for $(x, \xi)\in \mcw^+$,  
\begin{enumerate}
\item $(x, \xi)\in \WF(\tf_\times)  \text{ if and only if } (x, \xi)\in \WF(H_\times Xu)$. 
\item $(x, \xi)\in \WF(\tf_+)  \text{ if and only if } (x, \xi)\in \WF(H_+ Xu).$ 
\end{enumerate}
Also, there exists operators $H_\times', H_+'$ given in \eqref{eq-htimes} and \eqref{eq-hplus} such that for $(x, \xi)\in \mcw^-$,  
\begin{enumerate} 
\item $(x, \xi)\in \WF(\th_\times) \text{ if and only if } (x, \xi)\in \WF(H_\times' Xu)$. 
\item $(x, \xi)\in \WF(\th_+) \text{ if and only if } (x, \xi)\in \WF(H_+' Xu).$
\end{enumerate}
\end{theorem}

We remark that the operators $H_\times, H_+, H_\times', H_+'$ can be found rather explicitly in the proof. Thus the theorem provides a way to identify the polarization modes from computing $H_\bullet Xu, H_\bullet'Xu, \bullet = +, \times.$ 

Finally, for certain $f, h$ (for example $h=0$), Theorem \ref{thm-main} can recover the full wave front set of $f, h$ on $\mcr = \mcr^+\cup \mcr^-$. Combining with Theorem \ref{thm-main0}, we can obtain a stable inversion result. Below, we let $\chi_\mcr$ be a smooth function on $\mbr^3$ supported in $\mcr$. 
\begin{theorem}\label{thm-main1}
Let $u$ be the solution of \eqref{eq-wave} with Cauchy data 
\beq
u(0, x) = \chi_\mcr(x)f(x), \quad \p_t u(0, x) = 0, 
\eeq
where $f \in H^s_{\loc}(\mbr^3), s>2$ is of the form \eqref{eq-fh}. Then there exists constant $C>0$ such that 
\beqq\label{eq-est}
\|\chi_\mcr f\|_{H^s(\mbr^3)}   \leq C\|Xu\|_{H^{s+1}(\mcc)}. 
\eeqq 
\end{theorem}
We remark that $u$ in Theorem \ref{thm-main1} may not be a TT wave because (ii) of the TT gauge conditions may not hold. Nevertheless, we can think of $u$ as a ``piece" of TT wave originating from $\mcr$. The estimate \eqref{eq-est} indicates that inverting $X$ for such $u$ is stable. This type of result is instructive for developing numerical reconstruction method, see for example \cite{ChWa}. In particular, one can treat the inversion of $X$ as a PDE constrained optimization problem. For $u$ in Theorem \ref{thm-main1} without $\chi_\mcr$, inversion of $X$ is expected to be severely ill-posed. It is common to incorporate   regularization terms in the optimization problem to deal with the missing information. But \eqref{eq-est} suggests that reconstructing $f$ on $\mcr$ is stable.  We refer to the numerical study in \cite{ChWa} for more details. 

\subsection{The outline and main idea}
We briefly discuss the proof and some related results in the literature. The transform $X$ in \eqref{eq-light} is the analogue of the Radon transform in X-ray tomography or geodesic ray transform on Riemannian manifolds, see for example \cite{Sha}. The difference is that only integrals along null geodesics are used in $X$ instead of all geodesics. Here, we focus on the tensor problem and refer to the introduction of \cite{ChWa} for results on the scalar problem. 
To invert the transform on tensors, the first difficulty is that, similarly to the Riemannian problem, the transform has a non-trivial null space. The characterization of the null space for general Lorentzian metrics is largely open. There are a few results for the Minkowski metric \cite{LOSU} and certain stationary and static space-times \cite{FIO, OPS}. In Section \ref{sec-uni}, we show directly that TT waves are not in the null space which gives Theorem \ref{thm-main0}.

The more serious difficulty is that even for tensors in the complement of the null space, not all information can be stably recovered from the transform. For the Minkowski space-time, the microlocal structure of $X$ was studied in \cite{LOSU} with applications to the search of cosmic strings. It is known that the transform is microlocally invertible in space-like directions, and fails to be invertible in time-like directions. The invertibility is subtle near the light-like directions which are associated with gravitational waves, see \cite{Wan1}. We refer to \cite{COW} for a numerical demonstration of these phenomena.   An important observation was made in \cite{VaWa} that by incorporating the evolution equation \eqref{eq-wave}, inversion of $X$ can be stabilized. This was done for the scalar problem  via some ``back-projection" which takes $Xu(x, v)$ on $\mbr^3\times \mbs^2$ to a function on $\mbr^3.$ In particular, one can integrate out the $v$ variable and show that 
\beqq\label{eq-iu}
\tilde I Xu(x) = \int_{\mbs^2} Xu(x, v)dv
\eeqq
can be decomposed into a sum of FIOs acting on $f, h$ on $\mbr^3$ which can be used to solve for $f, h$ microlocally. Unfortunately, this method does not work for the tensor problem, because the resulting operators in \eqref{eq-iu} have large microlocal kernels on tensors hence very little information can be recovered. The novelty of the paper is to  demonstrate several new back-projections adapted for TT waves, the combination of which allows us to invert $X$. 

The paper is organized as follows. We prove Theorem \ref{thm-main0} in Section \ref{sec-uni}. Then we analyze the back-projections in Section \ref{sec-back}. We recover the $\times$ polarizations in Section \ref{sec-times} and the $+$ polarizations in Section \ref{sec-plus}. Finally, we prove Theorem \ref{thm-main} and Theorem \ref{thm-main1} in Section \ref{sec-pf}.

\section{The unique determination}\label{sec-uni}
We prove Theorem \ref{thm-main0} in this section. We consider the solution $u$ of \eqref{eq-wave} with Cauchy data of the form \eqref{eq-fh}. Because the equations are decoupled, we  know that $u$ is of the form 
\beqq\label{eq-u}
u(t, x) = 
\begin{pmatrix}
0 & 0 & 0 & 0 \\
0 & u_+(t, x) & u_\times(t, x) & 0 \\
0 & u_\times(t, x) & -u_+(t, x) & 0 \\
0 & 0 & 0 & 0 
\end{pmatrix}.  
\eeqq 
For $\eps>0$ small, we let $\chi_\eps\geq 0$ be a smooth cut-off function such that $\chi_\eps = 1$ on $\mcb_{1-\eps}$ and $\chi_\eps = 0$ outside of $\mcb_{1-\eps/2}$. Then we have $\chi_\eps Xu  = 0$ on $\mcc$. We can extend $\chi_\eps Xu$ to be identically zero on  $\mbr^3\times \mbs^2$. Taking the Fourier transform of $\chi_\eps Xu(y, \xi)$ in $y\in \mbr^3$, we get 
\beqq\label{eq-A}
\begin{split}
 0 &=   \int_0^T \int_{\mbr^3} e^{-iy\cdot \xi} \sum_{i, j = 0}^3 \chi_\eps(y) u_{ij}(t, y + t v - Tv)\theta^i\theta^j dy dt\\
 & = \int_\mbr \int_{\mbr^3} e^{-i(z - tv + Tv)\cdot \xi} \sum_{i, j = 0}^3 \chi_\dag(t) \chi_\eps(z-tv+ Tv) u_{ij}(t, z)\theta^i\theta^j dz dt\\
 & =   e^{-i Tv \cdot \xi} \sum_{i, j = 0}^3 \hat A_{ij}(-v\cdot \xi, \xi) \theta^i\theta^j, 
\end{split} 
\eeqq 
where we used $\chi_\dag(t)$ for the characteristic function of $[0, T]$ in $\mbr$, and $\hat A_{ij}(\tau, \xi)$ is the Fourier transform of 
\beq
A_{ij}(t, z) = \chi_\dag(t) \chi_\eps(z-tv+ Tv) u_{ij}(t, z), \quad i, j = 0, 1, 2, 3
\eeq 
for fixed $v\in \mbs^2$. Note that $A_{ij}(t, z)$ is compactly supported in $\mbr^4$ so $\hat A_{ij}$ is analytic in both $\tau$ and $\xi$. From \eqref{eq-A}, we see that $ \sum_{i, j = 0}^3 \hat A_{ij}(-v\cdot \xi, \xi) \theta^i\theta^j  = 0$ for any $v\in \mbs^2, \xi\in \mbr^3.$ Also, using \eqref{eq-u}, we get that only $\hat A_{11}( = -\hat A_{22}), \hat A_{12}( = \hat A_{21})$ are possibly non-zero and they satisfy  
\beqq\label{eq-A1}
\hat A_{11}(-v\cdot \xi, \xi)  v^1v^1 - \hat A_{11}(-v\cdot \xi, \xi) v^2v^2 + 2\hat A_{12}(-v\cdot \xi, \xi)  v^1v^2 = 0
\eeqq 
for all $v\in \mbs^2, \xi\in \mbr^3.$

Next, we choose $v$ to determine $f_{\times}, h_\times$. Let $a\in (0, \sqrt 2/2)$ and consider $v_a = (a, a, (1-2a^2)^\ha)\in \mbs^2$. It follows from \eqref{eq-A1} that 
\beq
\hat A_{12}(-v_a \cdot \xi, \xi) = 0. 
\eeq
Consider $\xi = (\xi_1, \xi_2, \xi_3)$ for $\xi_i \in \mbr, i = 1, 2, 3$. Then $v_a \cdot \xi =  a(\xi_1 + \xi_2) + (1-2a^2)^\ha \xi_3$. For any $\tau\in \mbr$, we see that $v_a\cdot \xi = \tau$ on a measure zero set of $\xi$. By the analyticity in $\xi,$ we see that $\hat A_{12}(-\tau, \xi)$ vanishes for all $\xi\in \mbr^3$. By varying $\tau,$ we conclude that $\hat A_{12}(\tau, \xi)$ must be identically zero. Thus, 
\beqq\label{eq-u12}
\chi_\dag(t) \chi_\eps(z-tv_a+ Tv_a) u_{12}(t, z) = 0
\eeqq
 for each $v_a$. When $t = 0$, we get $\chi_\eps(z + Tv_a) f_{\times}(z) = 0$. Note that $f_\times$ is a function of only $x_3$ variable. By the choice of $v_a$, we conclude that  $f_{\times} = 0$ in the set $\mcr_{T-1+\eps, T+1-\eps}.$ Because $\eps>0$ is arbitrary, we get that $f_{\times} = 0$ in the set $\mcr_{T-1, T+1}.$ By considering the $t$ derivative of \eqref{eq-u12}, we also get $h_{\times} = 0$ in $\mcr_{T-1, T+1}$.

Next, to determine $f_+, h_+,$ we set $w_a = (a, 0, (1-a^2)^\ha)$ where $a\in (0, 1)$. It follows from \eqref{eq-A1} that 
\beq
\hat A_{11}(-w_a\cdot \xi, \xi) = 0
\eeq
for all $\xi\in \mbr^3.$ For $\xi \in \mbr^3$, we have $w_a\cdot \xi =  a\xi_1 + (1-a^2)^\ha \xi_3$. 
We can repeat the argument for $\hat A_{12}$ to deduce that $\hat A_{11}$ must be identically zero for each $w$ in the family. Then we can follow the rest of the arguments for determining $f_\times, h_\times$ to conclude that $f_+ = h_+ = 0$ on $\mcr_{T-1, T+1}$. This completes the proof of Theorem \ref{thm-main0}.

\section{The analysis of the back-projection}\label{sec-back}
We start with some preparations for inverting $X$. Because $f, h$ in \eqref{eq-fh} only depend on $x_3$, it is possible to reduce \eqref{eq-wave} to a 1D wave equation and  find the d'Alembert solution. However,  the  oscillatory integral representation of the solution is more convenient for the analysis.  As we explained in Section \ref{sec-main}, $f, h$ outside of $\mcb_{T+1}$ does not contribute to $Xu.$ Let $\chi$ be a smooth cut-off function on $\mbr^3$ such that $\chi = 1$ on $\mcb_{T+1}$ and $\chi = 0$ outside of $\mcb_{T+2}$. By the principle of superposition for linear wave equations, we can decompose $u$ in \eqref{eq-wave} to $u = u_{in}+ u_{out}$ where 
\beqq\label{eq-wave1}
\begin{gathered}
\square u_{in}(t, x) = 0, \quad (t, x)\in \mcm \\
u_{in}(0, x) = \chi(x) f(x), \quad \p_t u_{in}(0, x) =  \chi(x) h(x), 
\end{gathered}
\eeqq
and 
\beqq\label{eq-wave2}
\begin{gathered}
\square u_{out}(t, x) = 0, \quad (t, x)\in \mcm \\
u_{out}(0, x) = (1- \chi(x)) f(x), \quad \p_t u_{out}(0, x) =  (1- \chi(x)) h(x). 
\end{gathered}
\eeqq
By the finite speed of propagation, we see that $Xu_{out} = 0$ on $\mcc$. For proving Theorem \ref{thm-main}, it suffices to take $u$ as $u_{in}$ and consider \eqref{eq-wave1}. The solution can be found by using Fourier transform. Let $\tf,  \th$ be the re-parametrized Cauchy data  
 \beqq\label{eq-fh0}
 \begin{gathered}
 \tf =   \chi f + \lap^{-\ha}   \chi h, \ \  \th =   \chi f  - \lap^{-\ha}    \chi h. 
\end{gathered}
 \eeqq
Here, $\lap^{-\ha}$ is defined by a Fourier multiplier of $|\xi|^{-1}$. 
Note that the equations in \eqref{eq-wave1} are decoupled. We find that 
\beqq\label{eq-cauchysol}
 u_{in; ij}  =  \ha (E^+  \tf_{ij} + E^-  \th_{ij}), \quad i, j  = 0, 1, 2, 3, 
 \eeqq
 where 
\beqq\label{eq-cauchypara1}
\begin{gathered}
(E^+  \tf_{ij})(t, x) = (2\pi)^{-3}\int_{\mbr^3}\int_{\mbr^3} e^{\imath ((x-y)\cdot \xi + t|\xi|)}   \tf_{ij}(y) dy d\xi,\\
(E^-  \th_{ij})(t, x) = (2\pi)^{-3}\int_{\mbr^3}\int_{\mbr^3} e^{\imath ((x-y)\cdot \xi -  t|\xi|)}   \th_{ij}(y) dy d\xi. 
\end{gathered}
\eeqq 
In fact, $E^\pm \in I^{-\frac 14}(\mbr^4,  \mbr^3; C^\pm)$ are Fourier integral operators associated with the canonical relations 
\beq
\begin{gathered}
C^\pm = \{( t, x, \tau, \xi; y, \eta)  \in T^*\mbr^4 \backslash 0 \times T^*\mbr^3 \backslash 0: \xi = \eta, x = y \pm t \eta/|\eta|, \\
\iota =  \tau = \pm |\eta|,  t\in (0, \infty), y\in \mbr^3, \eta\in \mbr^3\}.
\end{gathered}
\eeq   

Because of the special form \eqref{eq-fh} of $f, h$, we can write $\tf, \th$ defined by \eqref{eq-fh0}  in the form 
\beqq\label{eq-fh1}
\tf(x) = 
\begin{pmatrix}
0 & 0 & 0 & 0 \\
0 & \tf_+(x) & \tf_\times(x) & 0 \\
0 & \tf_\times(x) & -\tf_+(x) & 0 \\
0 & 0 & 0 & 0 
\end{pmatrix} , \quad 
\th(x) = 
\begin{pmatrix}
0 & 0 & 0 & 0 \\
0 & \th_+(x) & \th_\times(x) & 0 \\
0 & \th_\times(x) & -\th_+(x) & 0 \\
0 & 0 & 0 & 0 
\end{pmatrix}.
\eeqq
Also, we note that $\WF(f), \WF(h)\subset \mcw_0 \doteq \{(x, \xi)\in T^*\mbr^3\backslash 0: \xi = \la e_3, \la\in \mbr\}$. Because $\lap^{-1/2}$ can be regarded as a pseudo-differential operator of order $-1$, we see that $\WF(\tf), \WF(\th) \subset \mcw_0$. For $\eps>0$ small, we let $\phi_\eps(\xi)$ be a smooth cut-off function supported in a conic set 
\beqq\label{eq-veps}
\mcv_\eps = \{\xi\in \mbr^3\backslash 0: |\pm \xi/|\xi| - e_3|< \eps\}
\eeqq
so that $\phi_\eps \geq 0$ and $\phi_\eps(\xi) = 1$ on $\mcv_{\eps/2}\cap \{\xi\in \mbr^3: |\xi|>1/\eps\}$. We then split 
\beqq\label{eq-cauchypara2}
\begin{gathered}
E^\pm  = E^\pm_s  + E^\pm_r  , \text{ where }\\
E^\pm_s \psi(x) = (2\pi)^{-3}\int_{\mbr^3}\int_{\mbr^3} e^{\imath ((x-y)\cdot \xi \pm t|\xi|)} \phi_\eps(\xi)  \psi(y) dy d\xi,\\
E^\pm_r \psi(x) = (2\pi)^{-3}\int_{\mbr^3}\int_{\mbr^3} e^{\imath ((x-y)\cdot \xi \pm t|\xi|)} (1-\phi_\eps(\xi)) \psi(y) dy d\xi.  
\end{gathered}
\eeqq 
We observe that $E^\pm_r \psi \in C^\infty(\mcm)$ if $\WF(\psi)\subset \mcw_0.$ Using these expressions in \eqref{eq-cauchysol}, we have that  
\beqq\label{eq-xu}
\begin{split}
 Xu_{in} &= \ha X(E^+_s \tf + E^-_s \th) + R^+\tf + R^-\th   \\
&= \ha \sum_{i, j = 0}^3 (X^{ij} E^+_s \tf_{ij} + X^{ij} E^-_s \th_{ij}) + R^+\tf + R^-\th, 
\end{split}
\eeqq
where $R^\pm$ are regularizing operators, and  $X^{ij} \psi(x, v) = \tilde X \psi(x, v) \theta^i\theta^j, i, j = 0, 1, 2, 3$ where $\theta = (1, v), v\in \mbs^2$, and 
\beqq\label{eq-txu}
 \tilde X\psi(x, v) = \int_0^T \psi(t,   x + t v - Tv)  dt. 
\eeqq 
Here, $\tilde X$ is the  light-ray transform acting on scalar functions.  As noted in \cite{ChWa}, it is convenient to use a subset of the light rays for inverting $\tilde X$. For $\eps>0$ small, 
we use the light rays that meet $\{T+\eps\}\times \mcb_{1-\eps}$. 
For a scalar function $\psi$, we set  
\beqq\label{eq-lraye}
\tilde X_\eps \psi(x, v) = \int_0^T \chi_\eps(x) \psi(t, x + tv - (T +\eps)v)  dt. 
\eeqq
Note that when $t = T$, we have $x+tv-(T+\eps)v = x - \eps v \in \mcb_1$ for $x \in \mcb_{1-\eps}.$ The point is that we can think of $\tilde X_\eps \psi(x, v)$ as a function on $\mbr^3\times \mbs^2$ and the cut-off singularity at $\p \mcb_1$ in the $y$ variable is avoided. In the follows, we also use $X_\eps = \chi_\eps X.$ 

To recover information of $\tf, \th$ from $Xu$, we will use some ``back-projection" that takes functions on $\mbr^3\times \mbs^2$ to $\mbr^3.$ 
We start with the recovery of $\tf_\times, \th_\times.$ 
Let $a\in (0, \sqrt 2/2)$. Consider a family of unit vectors in $\mbr^3$ as $v_a = (a, a, (1-2a^2)^\ha)$. Using \eqref{eq-fh1}, we find that 
\beq
X_\eps u_{in}(x, v_a) =   2a^2  \tilde X_\eps E^+_s \tf_{\times}(x, v_a) +  2a^2  \tilde X_\eps E^-_s \th_{\times}(x, v_a) + R^+\tf + R^-\th.
\eeq
Let $\chi_*(a)$ be a smooth function for $a\in (0, \sqrt 2/2)$ such that $\chi_*\geq 0$, $\chi_*= 1$ on $(\sqrt 2/4, \sqrt 2/3)$ and $\chi_*$ is supported in $(\sqrt 2/5, \sqrt 3/3)$. We consider our first back-projection as 
\beqq\label{eq-iau}
I_{\tv}  X_\eps u_{in}(x) = \int_0^{\sqrt 2/2} \chi_*(a) X_\eps u_{in}(x, v_a) da. 
\eeqq
This makes sense for sufficiently regular $u_{in}$. In fact, we recall that $f\in H^{s}_{\loc}(\mbr^3), h\in H^{s-1}_{\loc}(\mbr^3)$. For the Cauchy problem \eqref{eq-wave1}, we know that $u_{in} \in H^{s}(\mcm)$ and 
\beq
\|u_{in}\|_{H^s(\mcm)}\leq C(\|\chi f\|_{H^s(\mbr^3)} + \|\chi h\|_{H^{s-1}(\mbr^3)}). 
\eeq
For $s> 2$ and by the Sobolev embedding, we know that $u_{in}$ is continuous on $\mcm.$ Thus $X_\eps u_{in}(x, v)$ is also continuous and \eqref{eq-iau} makes sense. 

We compute   \eqref{eq-iau} using \eqref{eq-xu}. First, we have 
\beq
\begin{gathered}
I_{\tv} (2 a^2\tilde X_\eps E^+_s \tf_\times)(x)  
 = (2\pi)^{-3} \int_0^{\sqrt 2/2} \int_{0}^T\int_{\mbr^3}\int_{\mbr^3} e^{\imath(x - y)\cdot \xi} e^{\imath (t - T)v_a \cdot \xi} e^{\imath t |\xi|}  \\
 \cdot 2a^2 \chi_\ast(a) \chi_\eps(x) \phi_\eps(\xi)  \tf_\times(y) dy d\xi dt da.   
  \end{gathered}
  \eeq
  We compute the kernel of the operator in the sense of distributions. Integrating in $t$, we get that 
\beqq\label{eq-iauu1}
\begin{gathered}
I_{\tv}  (2 a^2 \tilde X_\eps  E^+_s  \tf_\times)(x)  = (2\pi)^{-3} \int_0^{\sqrt 2/2} \int_{\mbr^3}\int_{\mbr^3} e^{\imath(x - y)\cdot \xi} e^{\imath (- T)v_a \cdot \xi}  \frac{2a^2 \chi_\ast(a) }{\imath |\xi|(v_a \cdot \xi/|\xi| + 1)} \\
\cdot (e^{\imath T(v_a \cdot \xi + |\xi|)} - 1) \chi_\eps(x)\phi_\eps(\xi) \tf_\times(y) dy d\xi  da
  = A_1^+\tf_\times + A_2^+\tf_\times, 
  \end{gathered}
  \eeqq
  where 
  \beq
  \begin{gathered}
A_1^+\tf_\times(x) = (2\pi)^{-3} \int_0^{\sqrt 2/2} \int_{\mbr^3}\int_{\mbr^3}  e^{\imath(x - y)\cdot \xi} e^{\imath T|\xi|}  \frac{2a^2 \chi_*(a)}{\imath |\xi|(v_a \cdot \xi/|\xi| + 1)}   \chi_\eps(x) \phi_\eps(\xi) \tf_\times(y) dy d\xi da, \\
A_2^+\tf_\times(x) = (2\pi)^{-3} \int_0^{\sqrt 2/2}  \int_{\mbr^3}\int_{\mbr^3}  e^{\imath(x - y)\cdot \xi} e^{-\imath T  v_a \cdot \xi}  \frac{-2a^2 \chi_*(a)}{\imath |\xi|(v_a \cdot \xi/|\xi| + 1)}\chi_\eps(x)  \phi_\eps(\xi) \tf_\times(y) dy d\xi da.  
  \end{gathered}
  \eeq
 Now we integrate in $a$. First, we have 
 \beq
 A_1^+\tf_\times(x)  = (2\pi)^{-3}  \int_{\mbr^3}\int_{\mbr^3}  e^{\imath(x - y)\cdot \xi} e^{\imath T|\xi|} a_1^+(x, \xi)\tf_\times(y) dy d\xi,
 \eeq 
 where 
 \beqq\label{eq-sym-a1p}
a_1^+(x, \xi) = \int_0^{\sqrt 2/2}  \frac{2 a^2\chi_*(a)}{\imath |\xi|(v_a \cdot \xi/|\xi| + 1)}  \chi_\eps(x) \phi_\eps(\xi) da. 
 \eeqq
 We note that for $\xi$ in the support of $\phi_\eps(\xi)$ with $\eps>0$ sufficiently small, the integrand is integrable because $\xi/|\xi|$ is close to $\pm(0, 0, 1)$ so  $v_a\cdot \xi/|\xi|$ is close to $\pm(1 - 2a^2)^\ha $ and the denominator in the integrand of \eqref{eq-sym-a1p} is non-zero for $a$ on the support of $\chi_*$. Furthermore, $a_1^+$ is a standard symbol of order $-1$. Thus $A_1^+\in I^{-1}(\mbr^3, \mbr^3; C_{T})$ where for $t\in \mbr, $ we set
 \beqq\label{eq-ct}
 C_t = \{(x, \xi, y, \eta)\in T^*\mbr^3\backslash 0\times T^*\mbr^3\backslash 0: x = y - t\xi/|\xi|, \xi = \eta \}. 
 \eeqq
 Second, for $A_2^+$, we note that the phase function 
 \beq
 \Phi(a) = v_a \cdot \xi = a\xi_1 + a\xi_2 + (1 - 2a^2)^\ha \xi_3
 \eeq
 has no critical points for $a$ in the support of $\chi_\ast$ and $\xi$ in the support of $\phi_\eps$ for $\eps>0$ sufficiently small, because 
 \beq
 \Phi'(a) =  \xi_1 +  \xi_2  + \frac{-a}{(1 - 2 a^2)^\ha} \xi_3, 
 \eeq
which is non-zero for $\xi/|\xi|$ close to $(0, 0, 1)$ and $a> \sqrt 2/5.$ Thus the operator $A_2^+$ is regularizing. 

For the other term in \eqref{eq-iau}, that is $I_\tv (2a^2 \tilde X_\eps E^-_s)$, the calculation  is similar. We have 
\beq
\begin{gathered}
\tilde X_\eps  E^-_s  \th_\times(x)
 = (2\pi)^{-3}   \int_{0}^T\int_{\mbr^3}\int_{\mbr^3}  e^{\imath(x - y)\cdot \xi} e^{\imath (t - T)v_a \cdot \xi} e^{-\imath t |\xi|} \chi_\eps(x)  \phi_\eps(\xi) \th_\times(y) dy d\xi dt.   
  \end{gathered}
  \eeq
Integrating in $t$, we get that 
\beqq\label{eq-iauu2}
\begin{gathered}
I_\tv( 2 a^2 \tilde X_\eps  E^-_s  \th_\times)(x) 
 = (2\pi)^{-3}  \int_0^{\sqrt 2/2} \int_{\mbr^3}\int_{\mbr^3}  e^{\imath(x - y)\cdot \xi} e^{-\imath T v_a \cdot \xi}  \frac{ 2a^2 \chi_*(a)}{\imath |\xi|(v_a \cdot \xi/|\xi| - 1)} \\
 \cdot (e^{\imath T(v_a \cdot \xi - |\xi|)} - 1)\chi_\eps(x) \phi_\eps(\xi) \th_\times(y) dy d\xi da  = A_1^-\th_\times + A_2^-\th_\times,
  \end{gathered}
  \eeqq
  where 
  \beq
  \begin{gathered}
  A_1^-\th_\times(x) = (2\pi)^{-3} \int_0^{\sqrt 2/2} \int_{\mbr^3}\int_{\mbr^3}  e^{\imath(x - y)\cdot \xi} e^{-\imath T|\xi|}  \frac{2a^2\chi_*(a)}{\imath |\xi|(v_a \cdot \xi/|\xi| - 1)}   \chi_\eps(x)  \phi_\eps(\xi) \th_\times(y) dy d\xi da, \\
  A_2^-\th_\times(x) = (2\pi)^{-3} \int_0^{\sqrt 2/2}  \int_{\mbr^3}\int_{\mbr^3}  e^{\imath(x - y)\cdot \xi} e^{-\imath T  v_a \cdot \xi}  \frac{-2a^2\chi_*(a)}{\imath |\xi|(v_a \cdot \xi/|\xi| - 1)} \chi_\eps(x)  \phi_\eps(\xi) \th_\times(y) dy d\xi da. 
  \end{gathered}
  \eeq
  Integrating in $a$, we find that  $A_2^-$ is regularizing and 
   \beq
A_1^-\th_\times(x)  = (2\pi)^{-3}  \int_{\mbr^3}\int_{\mbr^3}  e^{\imath(x - y)\cdot \xi} e^{-\imath T|\xi|} a_1^-(x, \xi)\th_\times(y) dy d\xi,
 \eeq 
 where 
 \beqq\label{eq-sym-a1m}
a_1^-(x, \xi) = \int_0^{\sqrt 2/2}   \frac{2 a^2\chi_*(a)}{\imath |\xi|(v_a \cdot \xi/|\xi| - 1)} \chi_\eps(x)   \phi_\eps(\xi) da.
 \eeqq
One can see that $a_1^-$ is also a symbol of order $-1$ so $A_1^-\in I^{-1}(\mbr^3, \mbr^3; C_{-T})$. 
  
Now we summarize our calculations \eqref{eq-iauu1} and \eqref{eq-iauu2} to get  
\beqq\label{eq-eqa}
I_\tv  X_\eps u_{in} =   A_1^+ \tf_{\times} +  A_1^- \th_{\times} + R^+\tf + R^-\th,
\eeqq 
with different regularizing operators $R^\pm$. Hereafter, we use $R^\pm$ for generic regularizing operators. To solve for $\tf_\times, \th_\times$, this equation alone is not enough. Below, we obtain another equation by varying the back-projection.

We choose another family of unit vectors in $\mbr^3$ as  $\bar v_a = (a, a, -(1 - 2a^2)^\ha)$, where $a\in (0,  \sqrt 2/2)$. Note that we flipped the sign of the last component compared to $v_a$. Using \eqref{eq-fh1}, we find that 
\beq
X_\eps u_{in}(x, \bar v_a) =   2a^2  \tilde X_\eps E^+_s \tf_{\times}(x, \bar v_a) +  2a^2  \tilde X_\eps E^-_s \th_{\times}(x, \bar v_a) +  R^+\tf + R^-\th.  
\eeq
We then consider the second back-projection as 
\beqq\label{eq-iau1}
I_{\bar \tv} X_\eps u_{in}(x) = \int_0^{\sqrt 2/2} \chi_*(a)  X_\eps u_{in}(x, \bar v_a)da.
\eeqq
We can use the calculation leading to \eqref{eq-eqa} by replacing $v_a$ by $\bar v_a$ to conclude that 
\beqq\label{eq-eqb}
I_{\bar \tv} X_\eps u_{in} =  B_1^+ \tf_{\times} + B_1^- \th_{\times} +  R^+\tf + R^-\th,
\eeqq
 where  
  \begin{enumerate}
 \item  $B_1^+\in I^{-1}(\mbr^3, \mbr^3; C_T)$ is given by 
    \beq
 B_1^+ \tf_\times(x)  = (2\pi)^{-3}  \int_{\mbr^3}\int_{\mbr^3}  e^{\imath(x - y)\cdot \xi} e^{\imath T|\xi|} b_1^+(x, \xi)\tf_\times(y) dy d\xi,
 \eeq 
where  
 \beqq\label{eq-sym-b1p}
b_1^+(x, \xi) = \int_0^{\sqrt 2/2} \frac{2 a^2\chi_*(a)}{\imath |\xi|(\bar v_a \cdot \xi/|\xi| +1)}\chi_\eps(x)   \phi_\eps(\xi) da, 
 \eeqq
\item $B_1^-\in I^{-1}(\mbr^3, \mbr^3; C_{-T})$ is given by 
   \beq
 B_1^-\th_\times(x)  = (2\pi)^{-3}  \int_{\mbr^3}\int_{\mbr^3}  e^{\imath(x - y)\cdot \xi} e^{-\imath T|\xi|} b_1^-(x, \xi)\th_\times(y) dy d\xi,
 \eeq 
where
 \beqq\label{eq-sym-b1m}
b_1^-(x, \xi) = \int_0^{\sqrt 2/2}  \frac{2 a^2\chi_*(a)}{\imath |\xi|(\bar v_a \cdot \xi/|\xi| - 1)} \chi_\eps(x)  \phi_\eps(\xi) da. 
 \eeqq
 \end{enumerate}
We remark that the choice of $\bar v_a$ does not change the types of FIOs in the decomposition of $I_{\bar \tv} X_\eps u$. It only changes the symbols.

 \section{The microlocal inversion for the $\times$ polarization}\label{sec-times}
We use \eqref{eq-eqa} and \eqref{eq-eqb} to microlocally solve for $\tf_\times, \th_\times.$ Note that $A_1^+, B_1^+ \in I^{-1}(\mbr^3, \mbr^3; C_{T})$ are FIOs of graph type, see \cite{Ho4}. The principal symbols are non-vanishing on the Lagrangian 
\beq
\La_{T} = \{(x, y, \xi, \eta)\in T^*(\mbr^3 \times \mbr^3) \backslash 0:  x = y - T\xi/|\xi|, \xi = -\eta, x \in \mcb_{1-\eps}, \xi\in \mcv_{\eps/2}\}. 
\eeq
In fact, the principal symbols  are given by 
\beq
\sigma_{1}(A_1^+)(x, y, \xi, \eta) =  a_1^+(x, \xi), \quad \sigma_{1}(B_1^+)(x, y, \xi, \eta) =  b_1^+(x, \xi)
\eeq  
multiplied by the half density factor and the Maslov factor. We will not show these factors because the only property we need is that they are non-zero. Note that the principal symbols are non-vanishing on $\La_T$. We can find parametrices $P^+, Q^+ \in I^{1}(\mbr^3, \mbr^3; C_{-T})$ such that 
 \beqq\label{eq-para}
\begin{gathered}
 P^+ A_1^+ \tf_\times = G_1^+ \tf_\times + R_1\tf_\times, \quad Q^+ B_1^+\tf_\times = G_1^+ \tf_\times + R_2\tf_\times, 
\end{gathered}
\eeqq
where $G_1^+$ is an pseudo-differential operator of order zero on $\mbr^3$ with symbol 
\beq
\sigma_0(G_1^+)(x, \xi) = \chi_{2\eps}(x-T\xi/|\xi|)\phi_{\eps/2}(\xi).
\eeq 
Also, $R_1, R_2$ are regularizing operators. The principal symbols of $P^+, Q^+$ on the Lagrangian $\La_{-T}$ are given by 
\beq
\sigma_{-1}(P^+)(x, y, \xi, \eta) = \chi_{2\eps}(y)\phi_{\eps/2}(\eta)/a_1^+(y, \eta), \quad \sigma_{-1}(Q^+)(x, y, \xi, \eta) =  \chi_{2\eps}(y)\phi_{\eps/2}(\eta)/b_1^+(y, \eta)
\eeq
multiplied by the half-density factor and Maslov factor. Using \eqref{eq-para}, we get 
 \beqq\label{eq-para-0}
 \begin{gathered}
G_1^+ \tf_\times + P^+ A_1^- \th_\times + R_1\tf_\times = P^+I_{\tv}  X_\eps u_{in} + R^+\tf + R^-\th,\\
G_1^+ \tf_\times + Q^+ B_1^-\th_\times + R_2\tf_\times = Q^+I_{\bar \tv}  X_\eps u_{in} + R^+\tf + R^-\th. 
 \end{gathered}
 \eeqq
We can absorb $R_1, R_2$ to $R^\pm$ and get that 
  \beqq\label{eq-para-1}
  (P^+ A_1^-   - Q^+ B_1^-)\th_\times  = P^+ I_\tv X_\eps u_{in} - Q^+ I_{\bar \tv}X_\eps u_{in} +  R^+\tf + R^-\th. 
  \eeqq
We know that $P^+A_1^-, Q^+B_1^- \in I^{0}(\mbr^3, \mbr^3; C_{-2T})$. We consider the principal symbol on the Lagrangian $\La_{-2T}$ at  $(x, \xi, y, \eta)$. Using the calculus of FIOs of the graph type, we get  
  \beq
  \begin{gathered}
  \sigma_0(P^+A_1^-)(x, \xi, y, \eta) = \sigma_{1}(P^+)(x, \xi, z, \zeta)\sigma_{-1}(A_1^-)(z, \zeta, y, \eta) =  \chi_{2\eps}(z)\phi_{\eps/2}(\zeta) a_1^-(z, \zeta)/a_1^+(z, \zeta), 
  \end{gathered}
  \eeq
  where $(z, \zeta) = (y - T\xi/|\xi|, \xi)$ with $z\in \mcb_{1-\eps}, \zeta\in \mcv_{\eps}$. Similarly, we have  
    \beq
  \sigma_0(Q^+ B_1^-)(x, \xi, y, \eta) = \sigma_{1}(Q^+)(x, \xi, z, \zeta)\sigma_{-1}(B_1^-)(z, \zeta, y, \eta) = \chi_{2\eps}(z)\phi_{\eps/2}(\zeta) b_1^-(z, \zeta)/b_1^+(z, \zeta). 
  \eeq
For $\zeta = \la e_3, \la\neq0$, using the symbol expressions \eqref{eq-sym-a1p}, \eqref{eq-sym-a1m}, \eqref{eq-sym-b1p} and \eqref{eq-sym-b1m}, we evaluate that 
\beq
\begin{gathered}
a_1^-(z, \zeta)/a_1^+(z, \zeta) = (\int_0^{\sqrt 2/2}   \frac{2 a^2\chi_*(a)}{(\frac{-a}{(1 - 2a^2)^\ha} -1)}da)/ ( \int_0^{\sqrt 2/2}  \frac{2 a^2\chi_*(a)}{(\frac{-a}{(1 - 2a^2)^\ha} + 1)} da), \\
b_1^-(z, \zeta)/b_1^+(z, \zeta) =   (\int_0^{\sqrt 2/2}   \frac{2 a^2\chi_*(a)}{(\frac{a}{(1 - 2a^2)^\ha} -1)}da)/ ( \int_0^{\sqrt 2/2}  \frac{2 a^2\chi_*(a)}{(\frac{a}{(1 - 2a^2)^\ha } + 1)} da). 
 \end{gathered}
\eeq
Now we claim that  $\sigma_0(P^+ A_1^- - Q^+ B_1^-)(x, \xi, y, \eta)\neq 0$. It suffices to show that 
\beq
\begin{gathered}
\int_0^{\sqrt 2/2}   \frac{2 a^2\chi_*(a)}{(\frac{a}{(1 - 2a^2)^\ha} +1)}da \neq \pm  \int_0^{\sqrt 2/2}  \frac{2 a^2\chi_*(a)}{(\frac{a}{(1 - 2a^2)^\ha} - 1)} da. 
 \end{gathered}
\eeq
We compute that 
\beq
\begin{gathered}
\int_0^{\sqrt 2/2}   \frac{2 a^2\chi_*(a)}{(\frac{a}{(1 - 2a^2)^\ha} +1)}da - \int_0^{\sqrt 2/2}  \frac{2 a^2\chi_*(a)}{(\frac{a}{(1 - 2a^2)^\ha} - 1)} da 
 = \int_0^{\sqrt 2/2} \frac{4a^2 (1-2a^2)}{1-3a^2}\chi_*(a)da.
 \end{gathered}
\eeq
Also, 
\beq
\begin{gathered}
\int_0^{\sqrt 2/2}   \frac{2 a^2\chi_*(a)}{(\frac{a}{(1 - 2a^2)^\ha} +1)}da + \int_0^{\sqrt 2/2}  \frac{2 a^2\chi_*(a)}{(\frac{a}{(1 - 2a^2)^\ha} - 1)} da 
 = \int_0^{\sqrt 2/2} \frac{4a^2(1-2a^2)^\ha}{3a^2 - 1}\chi_*(a)da.
 \end{gathered}
\eeq
Because $\chi_*(a)$ is supported on $(0, \sqrt 3/3)$, both terms are non-zero. Thus we proved that  $\sigma_0(P^+A_1^- - Q^+B_1^-)(x, \xi, y, \eta)\neq 0$. 

We can find a parametrix $W^+\in I^{0}(\mbr^3, \mbr^3; C_{2T})$ for $P^+A_1^- - Q^+ B_1^-$ so that 
\beq
W^+(P^+ A_1^- - Q^+ B_1^-)\th_\times = G_2^- \th_\times + R_3\th_\times, 
\eeq
where $R_3$ is regularizing and $G_2^-$ is a pseudo-differential operator of order zero on $\mbr^3$ with principal symbol
\beq
\sigma_0(G_2^-)(x, \xi) = \chi_{4\eps}(x+T\xi/|\xi|)\phi_{\eps/4}(\xi).
\eeq 
Using \eqref{eq-para-1}, we get 
\beqq\label{eq-para-2}
G_2^- \th_\times  = W^+ P^+ I_\tv  X_\eps u_{in} - W^+ Q^+ I_{\bar \tv}  X_\eps u_{in} + R^+\tf + R^-\th, 
\eeqq
where we absorbed $R_3\th_\times$ to $R^-\th.$ \\

To find $\tf_\times$, we can repeat the argument to first eliminate $\th_\times$. Starting from $A_1^-, B_1^- \in I^{-1}(\mbr^3, \mbr^3; C_{-T})$, we see that the principal symbols  on the Lagrangian $\La_{-T}$ are given by 
\beq
\sigma_{1}(A_1^-)(x, \xi, y, \eta) =  a_1^-(x, \xi), \quad \sigma_{1}(B_1^-)(x, \xi, y, \eta) =  b_1^-(x, \xi)
\eeq  
multiplied by the half density factor and the Maslov factor.  We can find parametrices $P^-, Q^- \in I^{1}(\mbr^3, \mbr^3; C_{T})$ such that 
 \beqq\label{eq-parapq}
\begin{gathered}
 P^-A_1^- \th_\times = G_1^- \th_\times + R_4\th_\times, \quad Q^- B_1^-\th_\times = G_1^- \th_\times + R_5\th_\times, 
\end{gathered}
\eeqq
where $G_1^-$ is an pseudo-differential operator of order zero on $\mbr^3$ with symbol 
\beq
\sigma_0(G_1^-)(x, \xi) = \chi_{2\eps}(x+T\xi/|\xi|)\phi_{\eps/2}(\xi).
\eeq 
and $R_4, R_5$ are regularizing operators. The principal symbols of $P^-, Q^-$ on the Lagrangian $\La_{T}$ are given by 
\beq
\sigma_{-1}(P^-)(x, \xi, y, \eta) = \chi_{2\eps}(x)\phi_{\eps/2}(\xi)/a_1^-(x, \xi), \quad \sigma_{-1}(Q^-)(x, \xi, y, \eta) =  \chi_{2\eps}(x)\phi_{\eps/2}(\xi)/b_1^-(x, \xi)
\eeq
multiplied by the half-density factor and Maslov factor. Using \eqref{eq-parapq}, we get 
 \beqq\label{eq-para-0-pq}
 \begin{gathered}
 P^-A_1^+ \tf_\times + G_1 \th_\times + R_4\th_\times = P^-I_{\tv}  X_\eps u_{in} + R^+\tf + R^-\th,\\
Q^-B_1^+ \tf_\times + G_1 \th_\times + R_5\th_\times = Q^-I_{\bar \tv}  X_\eps u_{in} + R^+\tf + R^-\th. 
 \end{gathered}
 \eeqq
We can absorb $R_4, R_5$ to $R^\pm$ and get that 
  \beqq\label{eq-para-1-pq}
  (P^- A_1^+   - Q^- B_1^+)\tf_\times  = P^- I_\tv X_\eps u_{in} - Q^- I_{\bar \tv}X_\eps u_{in} +  R^+\tf + R^-\th. 
  \eeqq
We know that $P^-A_1^+, Q^-B_1^+ \in I^{0}(\mbr^3, \mbr^3; C_{2T})$. We consider the principal symbol on the Lagrangian $\La_{2T}$ at  $(x, \xi, y, \eta)$. Using the calculus of FIOs of the graph type, we get  
  \beq
  \begin{gathered}
  \sigma_0(P^-A_1^+)(x, \xi, y, \eta) = \sigma_{1}(P^-)(x, \xi, z, \zeta)\sigma_{-1}(A_1^+)(z, \zeta, y, \eta) =  \chi_{2\eps}(z)\phi_{\eps/2}(\zeta)a_1^+(z, \zeta)/a_1^-(z, \zeta), 
  \end{gathered}
  \eeq
  where $(z, \zeta) = (y - T\xi/|\xi|, \xi)$. Similarly, we have  
    \beq
  \sigma_0(Q^- B_1^+)(x, \xi, y, \eta) = \sigma_{1}(Q^-)(x, \xi, z, \zeta)\sigma_{-1}(B_1^+)(z, \zeta, y, \eta) =   \chi_{2\eps}(z)\phi_{\eps/2}(\zeta) b_1^+(z, \zeta)/b_1^-(z, \zeta). 
  \eeq
By the same calculations, we conclude that  $\sigma_0(P^-A_1^+ - Q^-B_1^+)(x, \xi, y, \eta)\neq 0$. Now we can find a parametrix $W^- \in I^{0}(\mbr^3, \mbr^3; C_{-2T})$ so that 
\beq
W^-(P^+ A_1^- - Q^+ B_1^-)\tf_\times = G_2^+ \tf_\times + R_6\tf_\times, 
\eeq
where $R_6$ is regularizing and $G_2^+$ is a pseudo-differential operator of order zero on $\mbr^3$ with principal symbol
\beq
\sigma_0(G_2^+)(x, \xi) = \chi_{4\eps}(x-T\xi/|\xi|)\phi_{\eps/4}(\xi).
\eeq 
Using \eqref{eq-para-1-pq}, we get 
\beqq\label{eq-para-2-pq}
G_2^+ \tf_\times  = W^- P^- I_\tv  X_\eps u_{in} - W^- Q^- I_{\bar \tv}  X_\eps u_{in} + R^+\tf + R^-\th. 
\eeqq

\section{The analysis for the $+$ polarization}\label{sec-plus}
To recover $\tf_+, \th_+$, we essentailly repeat the argument for $\tf_\times, \th_\times$ with different choices of back-projections. This case is simpler and most of the calculations are similar. So we will not show all the details. 

First, for $a\in (0, \sqrt 2/2)$, we consider $w_a = (\sqrt 2 a, 0, (1-2a^2)^\ha)\in \mbs^2$. Then we note that 
\beq
 X_\eps u_{in} =  2a^2 \tilde X_\eps  E^+_s \tf_{+}  + R^+\tf + R^-\th. 
\eeq 
We consider 
\beq
I_\tw X_\eps u_{in}(x)  = \int_0^{\sqrt 2/2} \chi_\ast(a) X_\eps u_{in}(x, w_a)da
\eeq
and find that  
\beqq\label{eq-icu}
I_\tw  X_\eps u_{in}(x)  =  J_1^+ \tf_{+}(x) + R^+\tf + R^-\th, 
\eeqq
where $J_1^+\in I^{-1}(\mbr^3, \mbr^3; C_{T})$ is given by 
    \beq
    \begin{gathered}
J_1^+ \tf_+(x)  = (2\pi)^{-3}  \int_{\mbr^3}\int_{\mbr^3}  e^{\imath(x - y)\cdot \xi} e^{\imath T|\xi|} j_1^+(x, \xi)\tf_+(y) dy d\xi, \\
\text{ with } j_1^+(x, \xi) = \int_0^{\sqrt 2/2}  \frac{2a^2\chi_\ast(a)}{\imath |\xi|(w_a \cdot \xi/|\xi| +1)} \chi_\eps(x)  \phi_\eps(\xi) da. 
\end{gathered}
 \eeq 
From \eqref{eq-icu}, we can already solve for $\tf_+$. Note that $J_1^+ \in I^{-1}(\mbr^3, \mbr^3; C_{T})$ is an FIO of graph type. The principal symbol on the Lagrangian $\La_T$ is non-vanishing. We can find parametrices $U^+ \in I^{1}(\mbr^3, \mbr^3; C_{-T})$ such that 
\beqq\label{eq-para4}
U^+ J_1^+ \tf_+ = G_1^+ \tf_+ + R_7 \tf_+,
\eeqq
where $R_7$ is a regularizing operator. The principal symbols of $U^+$ on the Lagrangian $\La_{-T}$ is $\sigma_{-1}(U^+)(x, \xi, y, \eta) = \chi_{\eps/2}(x)\phi_{\eps/2}(\xi)/j_1^+(x, \xi)$ multiplied by the half-density factor and Maslov factor. Using \eqref{eq-para4}, we get 
 \beqq\label{eq-para-4}
 \begin{gathered}
 G_1^+ \tf_+  = U^+I_{\tw}  X_\eps u_{in} + R^+\tf + R^-\th. 
 \end{gathered}
 \eeqq

Second, we change the back-projection. We consider $\bar w_a = (0, \sqrt 2a, (1-2a^2)^\ha)$ for $a\in (0, \sqrt 2/2)$. Then 
\beq
X_\eps u_{in}  =  -2a^2  \tilde X_\eps   E^-_s \th_{+} + R^+\tf + R^-\th. 
\eeq 
We consider  
\beq
I_{\bar \tw} X_\eps u_{in}(x) = \int_0^{\sqrt 2/2} \chi_*(a) X_\eps u_{in}(x, \bar w_a) da 
\eeq
and find that 
\beqq\label{eq-idu}
I_{\bar \tw} X_\eps u_{in} =  K_1^- \th_{+}  + R^+\tf + R^-\th, 
\eeqq
where $K_1^-\in I^{-1}(\mbr^3, \mbr^3; C_{-T})$ is given by 
\beq
\begin{gathered}
K_1^-\th_+(x)  = (2\pi)^{-3}  \int_{\mbr^3}\int_{\mbr^3}  e^{\imath(x - y)\cdot \xi} e^{-\imath T|\xi|} k_1^-(x, \xi)\th_+(y) dy d\xi,\\
\text{ with } k_1^-(x, \xi) = \int_0^{\sqrt 2/2}  \frac{2a^2\chi_*(a)}{\imath |\xi|(\bar w_a \cdot \xi/|\xi| - 1)} \chi_\eps(a) \phi_\eps(\xi) da. 
\end{gathered}
 \eeq
We can  solve for $\th_+$ from \eqref{eq-idu}. Note that $K_1^+ \in I^{-1}(\mbr^3, \mbr^3; C_{-T})$ is an FIO of graph type. The principal symbol on the Lagrangian $\La_T$ is non-vanishing. We can find parametrices $U^- \in I^{1}(\mbr^3, \mbr^3; C_{T})$ such that 
\beqq\label{eq-para5}
U^- K_1^- \th_+ = G_1^- \th_+ + R_8 \th_+, 
\eeqq
where $R_8$ is a regularizing operator. The principal symbols of $U^-$ on the Lagrangian $\La_{T}$ is $\sigma_{-1}(U^-)(x, \xi, y, \eta) = \chi_{\eps/2}(x)\phi_{\eps/2}(\xi)/k_1^-(x, \xi)$ multiplied by the half-density factor and Maslov factor. Using \eqref{eq-para5}, we get 
 \beqq\label{eq-para-5}
 \begin{gathered}
 G_1^- \th_+  = U^-I_{\bar\tw}  X_\eps u_{in} + R^+\tf + R^-\th. 
 \end{gathered}
 \eeqq

\section{Proof of the theorems}\label{sec-pf}
\bpf[Proof of  Theorem \ref{thm-main}] The proof follows from the microlocal inversion results \eqref{eq-para-2}, \eqref{eq-para-2-pq}, \eqref{eq-para-4} and \eqref{eq-para-5}, which are 
\beqq\label{eq-sum}
\begin{split}
G_2^+ \tf_\times  &= H_\times X_\eps u_{in} + R^+\tf + R^-\th, \quad 
G_2^- \th_\times  = H_\times'  X_\eps u_{in} + R^+\tf + R^-\th, \\
 G_1^+ \tf_+ & = H_+ X_\eps u_{in} + R^+\tf + R^-\th, \quad 
 G_1^- \th_+  = H_+' X_\eps u_{in} + R^+\tf + R^-\th, 
\end{split}
\eeqq
where 
\beqq\label{eq-hplus}
H_+ =  U^+I_{\tw}, \quad  H_+' =  U^-I_{\bar\tw}, 
\eeqq
and  
\beqq\label{eq-htimes}
\begin{gathered}
H_\times =  W^- P^- I_\tv  - W^- Q^- I_{\bar \tv} ,\quad  H_\times' =  W^+ P^+ I_\tv    - W^+ Q^+ I_{\bar \tv}. 
\end{gathered}
\eeqq
The wavefront set statements in Theorem \ref{thm-main} can be read off  from \eqref{eq-sum} by the fact that $G_j^\pm, j = 1, 2$ 
 are pseudo-differential operators elliptic on $\mcw^\pm$, respectively.   Also, the result holds for any $\eps>0$. Thus we proved Theorem \ref{thm-main}. 
 \epf

\bpf[Proof of Theorem \ref{thm-main1}]
Suppose $h = 0$. We can use the relation \eqref{eq-fh} to solve for $f_\bullet, \bullet = +, \times.$  We have
\beqq\label{eq-inv-1}
\begin{gathered}
(G_1^+ + G_1^-) \chi  f_+ = \ha (G_1^+ \tf_+ + G_1^- \th_+) = \tilde H_+ X_\eps u + R^+ f, \\
\text{ where }  \tilde H_+ = \ha (U^+I_{\tw}   + U^-I_{\bar\tw}). 
\end{gathered}
\eeqq 
Also, 
\beqq\label{eq-inv-2}
\begin{gathered}
(G_2^+ + G_2^-) \chi f_\times = \ha (G_2^+ \tf_\times + G_2^- \th_\times) = \tilde H_\times  X_\eps u   + R^+f, \\
\text{ where } 
\tilde H_\times = \ha (W^- P^- I_\tv  - W^- Q^- I_{\bar \tv}   + W^+ P^+ I_\tv    - W^+ Q^+ I_{\bar \tv}). 
\end{gathered}
\eeqq

Now we can derive Sobolev estimates for recovering $f$. We replace $f_\bullet$ in \eqref{eq-inv-1} and \eqref{eq-inv-2} by $\chi_\mcr f_\bullet, \bullet = +, \times$. Let $G_j = G_j^+ + G_j^-, j = 1, 2$. Because $\chi_\mcr$ is smooth and $\WF(f)\subset \mcw_0$, we first observe that for any $\rho\in \mbr$, there is $C_\rho>0$ such that 
\beqq\label{eq-est-f}
\begin{gathered}
\|(1 - G_1)\chi_\mcr f_+\|_{H^s}\leq C_\rho \|\chi_\mcr f_+\|_{H^\rho}, \quad
\|(1 - G_2)\chi_\mcr f_\times\|_{H^s}\leq C_\rho \|\chi_\mcr f_\times\|_{H^\rho}.
\end{gathered}
\eeqq
Next, using \eqref{eq-inv-1} and \eqref{eq-est-f}, we have 
\beq
\begin{split}
\|\chi_\mcr f_+\|_{H^s} & \leq \|G_1 \chi_\mcr f_+\|_{H^s}+  \|(1 - G_1)\chi_\mcr f_+\|_{H^s} 
\leq C \| \tilde H_+  X_\eps u\|_{H^s} + C_\rho \|\chi_\mcr f\|_{H^\rho}\\
&\leq C \|(I_\tw  + I_{\bar \tw})X_\eps u\|_{H^{s+1}} + C_\rho \|\chi_\mcr f\|_{H^\rho}  \leq C \|X_\eps u\|_{H^{s+1}(\mcc)} + C_\rho \|\chi_\mcr f\|_{H^\rho}, 
\end{split}
\eeq
where in the second line we used that $U^+$ is an FIO of graph type of order $1$ so $U^+$ is continuous from $H^{s+1}(\mbr^3)$ to $H^{s}(\mbr^3)$, see \cite[Section 25.3]{Ho4}. Similarly, for $f_\times$, we have 
\beq
\begin{split}
\|\chi_\mcr f_\times\|_{H^s} & \leq \|G_2 \chi_\mcr f_\times\|_{H^s}+  \|(1 - G_2)\chi_\mcr f_\times \|_{H^s} 
\leq C \| \tilde H_\times  X_\eps u\|_{H^s} + C_\rho \|\chi_\mcr f\|_{H^\rho}\\
&\leq C \| I_\tv X_\eps u\|_{H^{s+1}}  + C\|  I_{\bar \tv} X_\eps u\|_{H^{s+1}} + C_\rho \|\chi_\mcr f\|_{H^\rho} \\
& \leq C \|X_\eps u\|_{H^{s+1}(\mcc)} + C_\rho \|\chi_\mcr f\|_{H^\rho}. 
\end{split}
\eeq
To summarize, we get that 
\beqq\label{eq-sobest}
\begin{gathered}
\|\chi_\mcr f \|_{H^s}  \leq C \|X_\eps u\|_{H^{s+1}(\mcc)} + C_\rho \|\chi_\mcr f\|_{H^\rho}. 
\end{gathered}
\eeqq
Finally, to remove the last  term, we first have a uniqueness result that if $Xu = 0$ then $\chi_\mcr f = 0$. This follows from the argument in Theorem \ref{thm-main0} and the fact that $\chi_\mcr f$ is continuous. Then we can use the arguments in Theorem 2.1 of \cite{ChWa}, see also Theorem 1.1 of \cite{VaWa} to remove the last term in \eqref{eq-sobest}. This completes the proof of Theorem \ref{thm-main1}. 
\epf


\end{document}